\documentclass[%
 reprint,
 superscriptaddress,
 amsmath,amssymb,
 aps,
 prd,
]{revtex4-2}

\usepackage{graphicx}
\usepackage{dcolumn}
\usepackage{bm}
\usepackage{booktabs}
\usepackage{hyphenat}
\usepackage{array}

\usepackage{wasysym}
\usepackage[separate-uncertainty,retain-explicit-plus,per-mode=symbol]{siunitx}
\usepackage{array,mathtools,amssymb,dcolumn}
\usepackage[below]{placeins}

\usepackage[dvipsnames, table]{xcolor}
\usepackage{tikz}
\usepackage{multirow}
\usepackage{afterpage}
\usepackage{lineno}
\usepackage{paralist}
\usepackage{listings}
\usepackage{array}
\usepackage{cancel}
\usepackage{xspace}

\usepackage{stmaryrd}
\usepackage[version=4]{mhchem}

\newcommand{\Geant}{\texttt{Geant4}}

\newcommand{\CERNRoot}{\texttt{ROOT}}

\newcommand{\RAT}{\texttt{RAT}}

\DeclareSIUnit\c{\mbox{$c$}}
\DeclareSIUnit\week{w}
\DeclareSIUnit\year{yr}
\DeclareSIUnit\yr{yr}
\DeclareSIUnit\yr{yr}
\DeclareSIUnit\standard{std}
\DeclareSIUnit\str{sr}
\DeclareSIUnit\ppm{ppm}
\DeclareSIUnit\ppb{ppb}
\DeclareSIUnit\ppt{ppt}
\DeclareSIUnit\pe{PE}
\DeclareSIUnit\spe{SPE}
\DeclareSIUnit\ev{events}
\DeclareSIUnit\hit{hit}
\DeclareSIUnit\hits{hits}
\DeclareSIUnit\bin{(\mbox{5-PE}~bin)}
\DeclareSIUnit\sgm{\mbox{$\sigma$}}
\DeclareSIUnit\rms{RMS}
\DeclareSIUnit\keVr{\mbox{keV$_{\rm nr}$}}
\DeclareSIUnit\keVee{\mbox{keV$_{\rm ee}$}}
\DeclareSIUnit\ph{photons}
\DeclareSIUnit\pm{PMT}
\DeclareSIUnit\inch{''}
\DeclareSIUnit\bit{bit}

\DeclareSIUnit\sample{S}
\DeclareSIUnit\barn{b}
\DeclareSIUnit\bara{bar}
\DeclareSIUnit\Curie{Ci}
\DeclareSIUnit{\msun}{\mbox{M$_\odot$}}
\DeclareSIUnit\mK{\milli\kelvin}
\DeclareSIUnit\micron{\micro\metre}
\DeclareSIUnit\liveday{\mbox{live-days}}
\DeclareSIUnit\tonneday{\mbox{tonne$\cdot$day}}
\DeclareSIUnit\days{\mbox{days}}

\usepackage[symbol*]{footmisc}
\DefineFNsymbolsTM{otherfnsymbols}{%
  \textdagger    \dagger
  \textasteriskcentered *
  \textbardbl    \|%
  \textparagraph \mathparagraph
  \textbullet \circ
  \textdaggerdbl \ddagger
}%
\setfnsymbol{otherfnsymbols}

\newcommand{\MV}{\mbox{MV}}

\newcommand{\PMT}{\mbox{PMT}}

\newcommand{\PMTs}{\mbox{PMTs}}

\begin{document}

\preprint{APS/123-QED}

\title{Measurement of the muon flux at SNOLAB using the DEAP-3600 experiment}

\newcommand{\Alberta}{Department of Physics, University of Alberta, Edmonton, Alberta, T6G 2R3, Canada}
\newcommand{\AstroCeNT}{AstroCeNT, Nicolaus Copernicus Astronomical Center, Polish Academy of Sciences, Rektorska 4, 00-614 Warsaw, Poland}
\newcommand{\BHSU}{School of Natural Sciences, Black Hills State University, Spearfish, SD 57799, USA}
\newcommand{\Cagliari}{Physics Department, Universit\`a degli Studi di Cagliari, Cagliari 09042, Italy}
\newcommand{\UCR}{Center for Experimental Cosmology \& Instrumentation, Department of Physics and Astronomy, University of California, Riverside, CA 92521, USA}
\newcommand{\CNL}{Canadian Nuclear Laboratories, Chalk River, Ontario, K0J 1J0, Canada}
\newcommand{\Carleton}{Department of Physics, Carleton University, Ottawa, Ontario, K1S 5B6, Canada}
\newcommand{\CIEMAT}{Centro de Investigaciones Energ\'eticas, Medioambientales y Tecnol\'ogicas, Madrid 28040, Spain}
\newcommand{\Houston}{Department of Physics, University of Houston, Houston, TX 77204, USA}
\newcommand{\INFNCagliari}{INFN Cagliari, Cagliari 09042, Italy}
\newcommand{\INFNNapoli}{INFN Napoli, Napoli 80126, Italy}
\newcommand{\Mainz}{Institut f\"ur Kernphysik, Johannes Gutenberg-Universit\"at Mainz, 55128 Mainz, Germany}
\newcommand{\LNGS}{INFN Laboratori Nazionali del Gran Sasso, Assergi (AQ) 67100, Italy}
\newcommand{\LU}{School of Natural Sciences, Laurentian University, Sudbury, Ontario, P3E 2C6, Canada}
\newcommand{\LBNL}{Nuclear Science Division, Lawrence Berkeley National Laboratory, Berkeley, CA 94720, USA}
\newcommand{\UNAM}{Instituto de F\'isica, Universidad Nacional Aut\'onoma de M\'exico, A.\,P.~20-364, Ciudad de M\'exico~01000, Mexico}
\newcommand{\Napoli}{Physics Department, Universit\`a degli Studi ``Federico II'' di Napoli, Napoli 80126, Italy}
\newcommand{\Capodimonte}{Astronomical Observatory of Capodimonte, Salita Moiariello 16, I-80131 Napoli, Italy}
\newcommand{\MEPhI}{National Research Nuclear University MEPhI, Moscow 115409, Russia}
\newcommand{\Oxford}{Department of Physics, University of Oxford, Oxford, OX1 3PU, United Kingdom}
\newcommand{\IIPA}{International Institute for Particle Astrophysics, Polish Academy of Sciences,  Bartycka 18, 00-716 Warsaw, Poland}
\newcommand{\Princeton}{Physics Department, Princeton University, Princeton, NJ 08544, USA}
\newcommand{\Queens}{Department of Physics, Engineering Physics, and Astronomy, Queen's University, Kingston, Ontario, K7L 3N6, Canada}
\newcommand{\RHUL}{Royal Holloway University London, Egham Hill, Egham, Surrey TW20 0EX, United Kingdom}
\newcommand{\RAL}{Rutherford Appleton Laboratory, Harwell Oxford, Didcot OX11 0QX, United Kingdom}
\newcommand{\SL}{SNOLAB, Lively, Ontario, P3Y 1M3, Canada}
\newcommand{\Sussex}{University of Sussex, Sussex House, Brighton, East Sussex BN1 9RH, United Kingdom}
\newcommand{\TRIUMF}{TRIUMF, Vancouver, British Columbia, V6T 2A3, Canada}
\newcommand{\TUD}{Institut f\"ur Kern und Teilchenphysik, Technische Universit\"at Dresden, 01069 Dresden, Germany}
\newcommand{\TUM}{Department of Physics, Technische Universit\"at M\"unchen, 80333 Munich, Germany}
\newcommand{\MI}{Arthur B. McDonald Canadian Astroparticle Physics Research Institute, Queen's University, Kingston, ON, K7L 3N6, Canada}
\newcommand{\UIB}{IAC3, Universitat de les Illes Balears, Cra.~de Valldemossa km 7.5, 07122 Palma, Spain}

\affiliation{\Alberta}
\affiliation{\AstroCeNT}
\affiliation{\BHSU}
\affiliation{\Cagliari}
\affiliation{\UCR}
\affiliation{\CNL}
\affiliation{\Carleton}
\affiliation{\CIEMAT}
\affiliation{\Houston}
\affiliation{\INFNCagliari}
\affiliation{\INFNNapoli}
\affiliation{\Mainz}
\affiliation{\LU}
\affiliation{\UNAM}
\affiliation{\Capodimonte}
\affiliation{\Napoli}
\affiliation{\MEPhI}
\affiliation{\Oxford}
\affiliation{\IIPA}
\affiliation{\Queens}
\affiliation{\RHUL}
\affiliation{\RAL}
\affiliation{\SL}
\affiliation{\TRIUMF}
\affiliation{\TUD}
\affiliation{\TUM}
\affiliation{\MI}

\author{P.~Adhikari}\affiliation{\Carleton}
\author{M.~Alp\'izar-Venegas}\affiliation{\UNAM}
\author{P.-A.~Amaudruz}\affiliation{\TRIUMF}
\author{D.\,J.~Auty}\affiliation{\Alberta}
\author{M.~Batygov}\affiliation{\LU}
\author{B.~Beltran}\affiliation{\Alberta}
\author{M.\,A.~Bigentini}\affiliation{\Carleton}\affiliation{\MI}
\author{C.\,E.~Bina}\affiliation{\Alberta}\affiliation{\MI}
\author{W.~Bonivento}\affiliation{\INFNCagliari}
\author{M.\,G.~Boulay}\affiliation{\Carleton}
% ...

\collaboration{DEAP Collaboration}\email{deap-papers@snolab.ca}\noaffiliation

\date{\today}

\begin{abstract}

A direct measurement of the muon flux at SNOLAB is performed using the DEAP-3600 experiment, located 2 km underground at SNOLAB near Sudbury, Canada.
Primarily designed for the direct detection of weakly interacting massive particles (WIMPs), a dark matter candidate, DEAP-3600 consists of an inner spherical acrylic vessel containing a liquid argon target; this vessel is enclosed within a steel shell which is submerged in an instrumented water tank, serving as a muon veto for the dark matter search. 
The muon flux measurement is performed using a cut-and-count analysis of events observed in the muon veto detector and of events which are coincident between the muon veto and the liquid argon target. The requirement that muons traverse both the water and liquid argon minimizes instrumental backgrounds and systematic uncertainties. Using data collected from November 2016 to March 2020, the muon flux is measured by this coincidence analysis to be  $(3.71 \pm 0.25_{\textrm{stat}} \pm 0.09_{\textrm{sys}}) \times 10^{-10}\, \mu/$cm$^2$/s. 
The standalone measurement using muon veto data only is compatible within uncertainties.
Both measurements agree with the previous result by the SNO experiment and with simulations carried out using the MUTE software. These results provide an important benchmark for future rare-event searches at the SNOLAB facility.

\end{abstract}

\maketitle


\section{Introduction}
\label{sec:Introduction}

Rare-event searches such as those conducted by dark matter and neutrino experiments require extremely low background rates to achieve high sensitivity. Underground laboratories are ideal locations to operate such experiments, as the overburden of rock provides shielding against cosmic-ray-induced particles~\cite{ianni_review_2017, palusova_natural_2020, abi2021prospects, pvevc2024muon, das2026rare, abreu2026cosmogenic, xia2026progress}.

To further shield these experiments from external radioactivity and to identify muons which penetrate the overburden of rock, some experiments rely on a water-based \v{C}erenkov radiation detector surrounding their main inner detector, which acts as a muon veto (MV)~\cite{an_muon_2015, calkins_prototyping_2015, agnes_veto_2016, angloher2024water, burlac2025early}. The \v{C}erenkov radiation signals produced by incident muons and their associated electromagnetic shower products are used to mitigate signals in the inner detector from both neutron-induced events and events due to the decay of short-lived radioisotopes~\cite{oconnell_muon-induced_1988, carson_neutron_2004, mei_muon-induced_2006, kamland_collaboration_production_2010, bellini_cosmogenic_2013, empl_fluka_2014, malgin_phenomenology_2017, aprile2025neutron}. The cosmic-ray muon flux reaching the detector can also be measured using the coincident signal observed by a muon veto detector and its corresponding inner detector.

The DEAP-3600 experiment is located at the Sudbury Neutrino Observatory Laboratory (SNOLAB), situated in the Creighton nickel mine near Sudbury, Canada.
A depth of approximately 2~km of rock provides a flat overburden corresponding to $(5.89\pm0.94)~\mathrm{km.w.e}$ \cite{SNO_muonflux} at normal incidence, which significantly suppresses the flux of cosmic-ray muons reaching the laboratory \cite{ford2012snolab, hall2020snolab, diamond2025community}. 

Here we present a measurement of the muon flux at SNOLAB by the DEAP-3600 experiment based on events observed in its water-based \v{C}erenkov radiation detector (known as the MV) and events coincident between the MV and the liquid argon (LAr) detector. 
Section~\ref{sec:deapExperiment} presents the DEAP-3600 experiment and specifically the MV in more detail, with a summary of data collection, run selection, detector livetime calculation and simulation methods.
Section~\ref{sec:MuonFluxAtSNOLAB} recalls the previous result by the Sudbury Neutrino Observatory (SNO), presents the muon intensity parametrization of Mei and Hime~\cite{mei_muon-induced_2006} and features an a priori theoretical estimate based on simulations carried out with the MUon inTensity codE (MUTE)~\cite{MUTE, MUTE_software} software. 
The measurement of the muon flux with standalone MV events is reported in Section~\ref{sec:WaterTankMuonFlux}, while Section~\ref{sec:CoincidenceMuonFlux} describes the measurement of the muon flux with events measured in coincidence between the MV and the LAr detector. 
Lastly, Section~\ref{sec:results} presents a comparison of muon flux measurements and a brief discussion of the results and uncertainties.


\section{DEAP-3600 Experiment}
\label{sec:deapExperiment}

DEAP-3600, located in Cube Hall at SNOLAB, is designed to directly detect dark matter by searching for nuclear recoil (NR) signals produced by weakly interacting massive particles (WIMPs) scattering on $^{40}$Ar nuclei. 
This experiment uses pulse-shape discrimination (PSD) to separate electronic-recoil (ER) signals from NR signals in the WIMP search region of interest~\cite{deap_psd_2021}.

DEAP-3600 comprises a LAr target mass of \mbox{(3269 $\pm$ 24)}~kg~\cite{adhikari2023precision}, housed within a 5~cm thick acrylic vessel (AV). An array of 255 Hamamatsu R5912-HQE photomultiplier tubes (PMTs) looks into the AV via 45~cm-long cylindrical acrylic light guides, which provide both thermal insulation from the LAr and neutron shielding between the PMTs and the LAr. The AV is encased in a stainless steel shell (SSS) which is submerged in a cylindrical water tank that acts as the MV. 
The MV detector is described in detail in Section~\ref{subsec:vetoInstrumentation}. Ref.~\cite{amaudruz_design_2019} provides a comprehensive description of the DEAP-3600 experiment.

\subsection{The muon veto detector}
\label{subsec:vetoInstrumentation}

\begin{figure}[t]
    \centering
    \includegraphics[width=0.99\columnwidth]{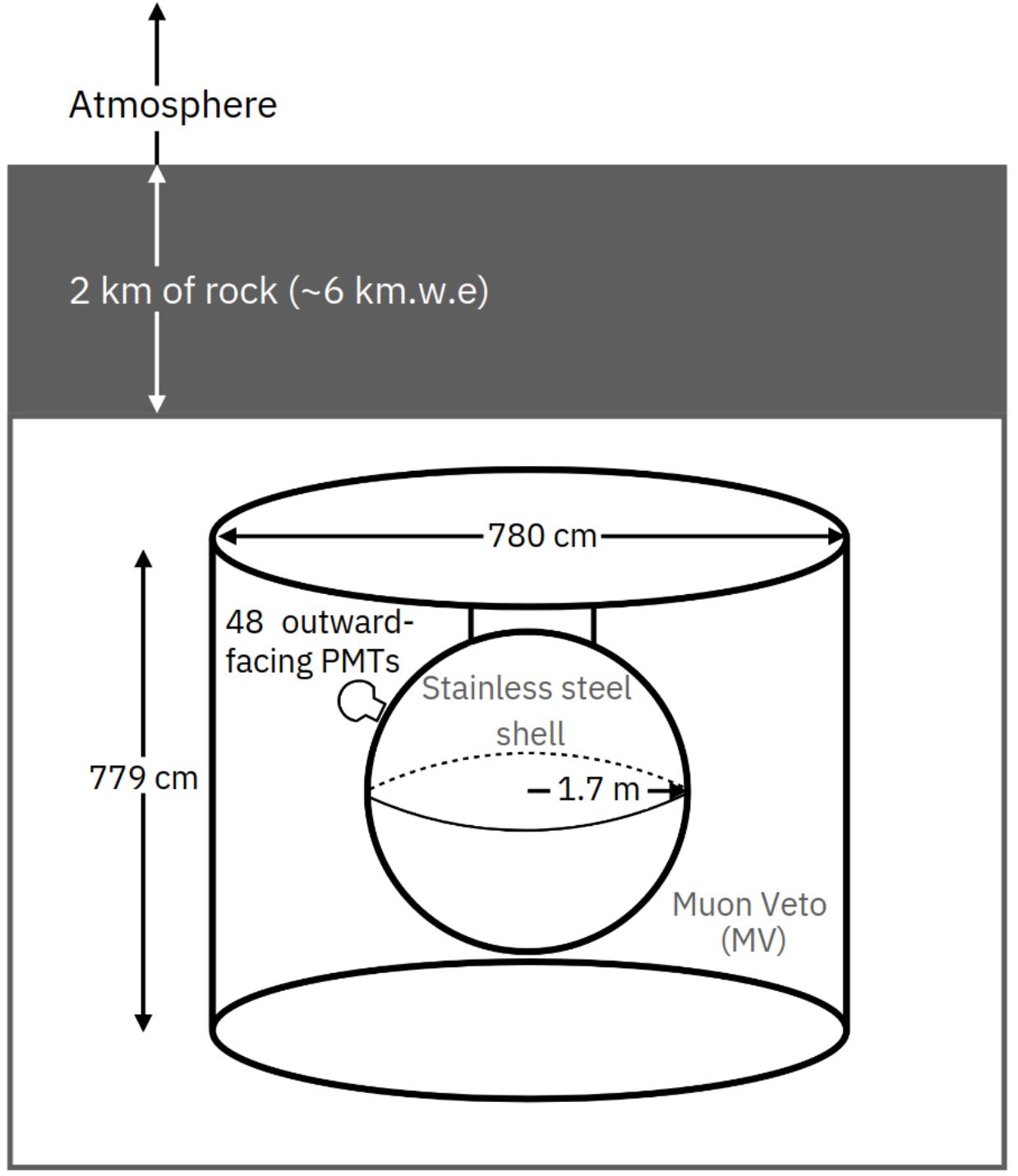}
    \caption{A schematic of the DEAP-3600 muon veto system (not to scale).}
    \label{fig:mvschem}
\end{figure}

The DEAP-3600 MV consists of a galvanized steel cylinder with a height of 779~cm and a diameter of 780~cm, filled with 373~tonnes of ultra-pure water (UPW). During the detector operation, the water level was maintained at $(751\pm8)$~cm. The exterior of the submerged SSS is instrumented with 48 outward-facing Hamamatsu R1408 PMTs. A schematic of this setup is shown in Fig.~\ref{fig:mvschem}. 

To detect muons and electromagnetic shower components that travel 
through the MV, high light-collection efficiency is required. To compensate for the small PMT coverage area, the inside of the water tank is covered with a vinyl liner made of PVC \#328 purchased from Kentain Products Limited. This liner is a white diffuse reflector, and its reflectivity was measured using a Perkin–Elmer Lambda 35 UV–Vis spectrometer with a Spectralon-coated integrating sphere (RSA-PE-20).

\begin{figure}[t]
    \centering
    \includegraphics[width=0.99\columnwidth]{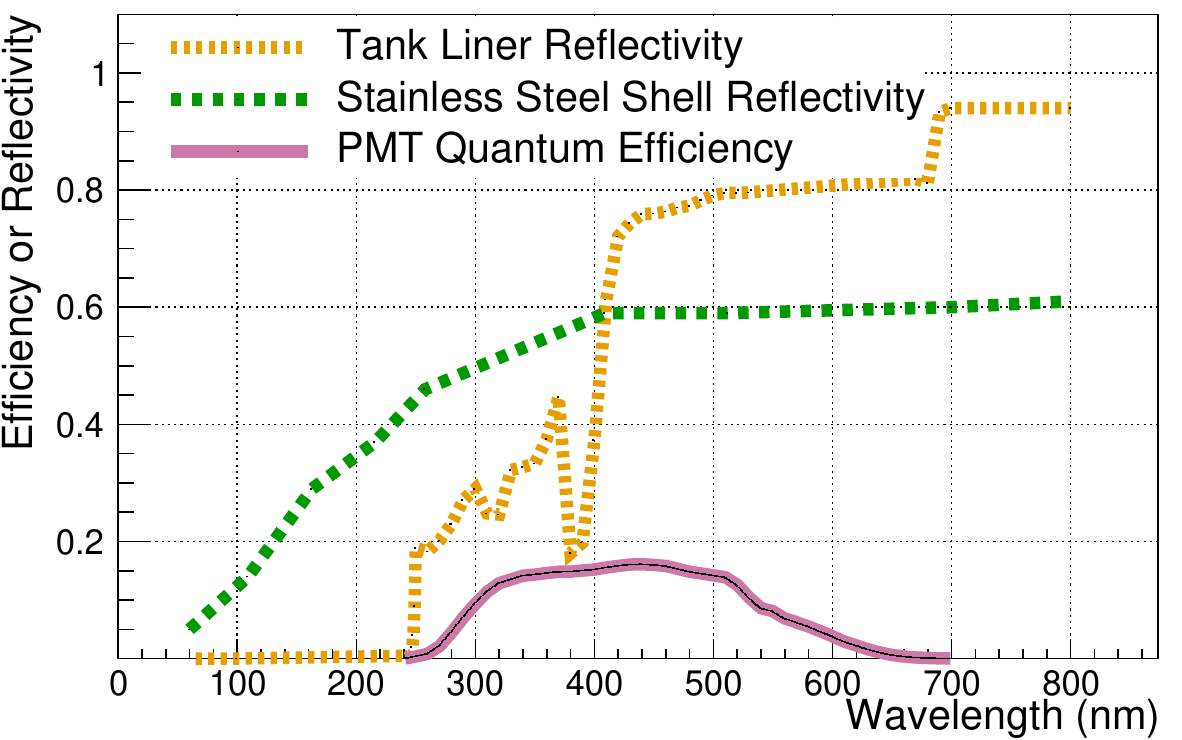}
    \caption{Quantum efficiency of the Hamamatsu R1480 PMTs as a function of wavelength, shown along with the coefficients of reflectivity for the water tank liner and stainless steel shell.}
    \vspace{0.25cm}
    \label{fig:qeandrefl}
\end{figure}

Optical photons can also reflect from the SSS. The reflectivity for this material is taken from Ref.~\cite{bass_handbook_2009}, where the complex index of refraction was computed using a molar-weighted average of the constituent elements. The reflectivity of the liner and the SSS are shown alongside the PMT quantum efficiency in Fig.~\ref{fig:qeandrefl}.

Because impurities in the water tank could absorb \v{C}erenkov photons, the water is continuously looped through a purification system to maintain high purity. This system consists of an Iwaki America MX-251 magnetically coupled pump, Purolite NRW-37 mixed-bed ion exchange columns, a Shelco 5FOS2 multi-cartridge filter housing containing 0.45~$\mu$m filters, and a Membrane Liqui-Cel which replaces oxygen in the water with pure nitrogen. The UPW is then sterilized using a 254~nm Viqua UV Max F4 Plus sterilizer. The vinyl liner also serves as a barrier to prevent the ingress of additional impurities while the MV is operating. The purity of the UPW is periodically monitored using CDCE-90-X Series Conductivity Sensors.

The water tank temperature is continuously monitored using a custom implementation of the \texttt{DeltaV} distributed control system.
From 2018, the temperature was kept stable at a nominal 284~K~\cite{AndrewThesis}
to ensure PMT stability and dark rate while keeping the outer surface of the water tank above the dew point at normal SNOLAB humidity.

Due to hardware configuration differences between the MV and the LAr detector, the reconstructed time of events observed in the MV is offset by approximately 4.43~$\mu$s with respect to the LAr detector; this is taken into account in the coincidence analysis.
Position reconstruction and tracking of muon candidate events from MV data is not attempted here.

\subsection{DEAP-3600 data}
\label{subsec:DEAP_data}

The data used in this analysis were collected between November 2016 and March 2020. Data were collected in discrete runs with a typical length of approximately 22~hours, though their length varied from less than 1~hour to up to 2~days.

Run stability and quality are assessed by examining \MV\ \PMT\ behaviour and trends in the MV observables. The selection criteria are based on runtime, PMT rates, PMT charge distributions, and MV observable characteristics. Runs of less than 8 hours are removed from the dataset to ensure sufficient statistics for analysis. 
The efficiencies and single-photoelectron charge response of individual PMTs are calibrated regularly as detailed in Refs.~\cite{ajaj2019search, AMAUDRUZ2019373}.
Runs with PMT rates that are not sufficiently constant are removed from the dataset. The PMT charge distributions are assessed by the fraction of events with charges between 30 and 50 photoelectrons and by the mean charge below 10 photoelectrons, and runs with PMT charges outside the nominal range are removed from the dataset. 

After all selection criteria are applied, a total of 376 runs remain. After applying a deadtime correction to each run, the total livetime of the dataset is $t_{\text{live}}^{\text{coinc}}~=~2.85~\times~10^7$~s. 
This dataset is used for the coincidence analysis.

The standalone water tank muon flux measurement requires an additional criterion for selecting runs based on the number of active MV PMTs. 
During the course of operation, some PMTs were deactivated due to either excessively high noise rates or electronic failures. As a result, starting in late 2019, the number of active PMTs slowly decreased from 48 to 38. 
Although inactive, the 10 deactivated PMTs introduced noise within the data acquisition system. 53 runs taken after MV PMTs began to fail are excluded from the standalone water tank muon flux measurement, resulting in a livetime for this analysis of $t_{\textrm{live}}^{\textrm{WT}}~=~2.44~\times~10^7$~s.
In contrast, in the coincidence analysis where muon candidates are identified by tagging events detected in both the MV and LAr detectors, this extra criterion is not required there, so the longer livetime given in the previous paragraph is used.

Event selection criteria are determined by comparing the observed data with Monte Carlo (MC) simulations of muons passing through DEAP-3600, generated with the \RAT\ software framework based on \CERNRoot~v5.34.36~\cite{ROOT_NIMA_1997} and \Geant~v9.6.2~\cite{Geant4_1, Geant4_2}. 
These event selection requirements enable the identification and removal of background events in the dataset, while preserving a high signal efficiency for muon events.
The event selection relevant to each analysis and the corresponding signal acceptance are detailed in Section~\ref{sec:WaterTankMuonFlux} for the standalone water tank measurement and in Section~\ref{sec:CoincidenceMuonFlux} for the coincidence measurement.


\section{Muon Flux at SNOLAB: \\ Previous Results}
\label{sec:MuonFluxAtSNOLAB}

\subsection{Sudbury Neutrino Observatory measurement}

In the measurement of the muon flux by SNO~\cite{SNO_muonflux}, 
the vertical muon intensity as a function of depth is parametrized as

\begin{equation}
 I(x_{\textrm{std}}) = I_0\, \left( \frac{x_0}{x_{\textrm{std}}} \right)^{\alpha} e^{-x_{\textrm{std}}/x_0}
 \label{eq:SNO_int_flux}
\end{equation}

\noindent where $I_0$ is the overall normalization constant, $x_0$ is the effective attenuation length for high-energy muons, $\alpha$ is a dimensionless power-law exponent, and $x_{\textrm{std}}$ is the equivalent slant depth in standard calcium carbonate rock derived by muon energy loss models, defined as

\begin{equation}
 x_{\textrm{std}} = 1.015\, x_{\text{SNO}} + \frac{x_{\text{SNO}}^2}{4 \times 10^5\, \textrm{m.w.e}} 
 \label{eq:SNO_vertical_depth}
\end{equation}

\noindent where $x_{\textrm{SNO}}$ is the slant depth based on the rock overburden and topography at SNOLAB. Ref.~\cite{SNO_muonflux} reports resulting parameter values of $I_0~=~(2.16 \pm 0.03)\times 10^{-6} \> \text{cm}^{-2} \> \text{s}^{-1} \> \text{sr}^{-1}$, $x_0~=~(1.14\pm~0.02) \> \text{km.w.e}$, and $\alpha=(1.87 \pm 0.06)$, estimated by fitting Eq. \ref{eq:SNO_int_flux} to data from LVD~\cite{aglietta1998muon}, MACRO~\cite{ambrosio1995vertical}, and SNO.

The SNO measurement of the cosmic-ray muon flux is

\begin{equation*}
 \Phi_{\text{SNO}} = (\>3.31 \pm 0.01_{\textrm{stat}} \pm 0.09_{\textrm{sys}}\>) \times 10^{-10}\> \mu/\textrm{cm}^2/\textrm{s}
 \label{eq:SNO_muon_flux}
\end{equation*}

\noindent where a requirement is applied to the reconstructed zenith angle: $\cos \theta_{\textrm{zenith}} > 0.4$.

\subsection{Mei and Hime model}

In the method described by Mei and Hime in Ref.~\cite{mei_muon-induced_2006} based on a depth-intensity relation by Groom \textit{et al.}, for an underground laboratory with a flat overburden, the through-going muon intensity can be expressed as a function of slant depth $h$ and zenith angle $\theta$ as 

\begin{equation}
  I_\textrm{th} (h, \theta) = \Big( I_1\, e^{-h/\lambda_1} + I_2\, e^{-h/\lambda_2}\Big)  \sec \theta
  \label{eq:mei-hime_int_flux_slant_depth_zenith}
\end{equation}

\noindent where $I_i$ are empirical normalization constants and $\lambda_i$ are characteristic attenuation lengths with values given in Ref.~\cite{mei_muon-induced_2006}. 
In this coordinate system, the muon flux is maximum at zero zenith angle.

In these variables, the SNO parametrization reads

\begin{equation}
 I_\textrm{th} (h, \theta) = I_0\, \left( \frac{x_0}{h} \right)^{\alpha} e^{-h / x_0} \sec \theta .
 \label{eq:SNO_int_flux_zenith_dependent}
\end{equation}

\noindent It is convenient to express the slant depth as a function of the vertical depth $h_0$ 
such that \mbox{$h = h_0 \sec \theta$},
here taking \mbox{$h_0 = 5.89$~km.w.e.}~\cite{SNO_muonflux}.

\subsection{Muon flux simulations with MUTE}

An estimate of the muon flux in the SNOLAB Cube Hall is performed using the open-source MUTE simulation software (v2.0.0), which is designed to calculate muon fluxes across various underground environments. MUTE was developed to provide a flexible and reproducible framework for calculating underground muon fluxes essential for cosmic-ray physics and rare-event searches. It consists of two main components: an efficient air shower simulation software called MCEq~\cite{MCEq} which calculates surface fluxes $\Phi^{s}$ by solving the one-dimensional cascade equation; a particle propagating tool called PROPOSAL~\cite{PROPOSAL} which propagates leptons and photons through rock and water and determines the underground flux $\Phi^{u}$ via the survival probability tensor $P$, as:

\begin{equation}
    \Phi^{u}(E^{u}_j, h_k, \theta_k) = \sum_i \Phi^{s}(E^{s}_j, \theta_k) P(E^{s}_i, E^{u}_j, h_k) \left(\frac{\Delta E^{s}_j}{\Delta E^{u}_j}\right)
    \label{eq:MUTE}
\end{equation}

\noindent where $E^u$ and $E^s$ are the muon energies underground and at the surface, respectively; $h$ is the slant depth, and $\theta$ is the zenith angle. The subscripts $i, j, k$ are indices over discrete bins for surface energy, underground energy, and zenith angle, respectively.

Within the simulation, the detector is situated at a vertical depth of 5.89~km.w.e with a flat rock overburden having a density of 2.83~g/cm$^{2}$. MUTE assumes a 2\% uncertainty on the rock density. For 100 energy-zenith bins, 1000 muons were generated, making a total of 10$^5$ muons. The flux is estimated for January, and the uncertainty in the seasonal variation is taken as approximately 3\%. 
The uncertainties from rock density and seasonal variation are negligible compared to those from the cosmic-ray spectrum and air-shower hadronic interactions, where disagreement exists among the various models.
Three prominent models of the cosmic-ray primary spectrum are considered: the Hillas and Gaisser (HG) model~\cite{H3a}, the Global Spline Fit (GSF) model~\cite{GSF}, and the simple power law (PL27)~\cite{thunman1996charm} model. Two hadronic models are considered for each CR spectrum: the SIBYLL 2.3c model~\cite{fedynitch2019hadronic} and the EPOS-LHC model~\cite{pierog2015epos}. The results of the simulations are shown in Tab.~\ref{tab:flux_MUTE}.

    \begin{table}[htb]
        \centering
        \renewcommand*{\arraystretch}{1.4}
        \caption{Muon flux at SNOLAB estimated with MUTE software for variations of the cosmic-ray primary spectrum model and the hadronic model considered.}
        \vspace{0.25cm}
        \begin{tabular}{c c c}
            \toprule
            \textbf{Cosmic-ray primary} & \textbf{Hadronic} & \textbf{Muon flux} \\
            \textbf{spectrum model} & \textbf{model} & \textbf{[$\times 10^{-10}$\,$\mu$/cm$^2$/s]} \\
            \midrule
            GSF & SIBYLL 2.3c & 4.18$^{+0.08}_{-0.02}$ \\
            GSF & EPOS-LHC & 4.34$^{+0.08}_{-0.02}$ \\
            HG & SIBYLL 2.3c & 4.62$^{+0.08}_{-0.02}$  \\
            HG & EPOS-LHC & 4.80$^{+0.08}_{-0.02}$ \\
            PL27 & SIBYLL 2.3c & 3.94$^{+0.07}_{-0.02}$  \\
            PL27 & EPOS-LHC & 4.07$^{+0.07}_{-0.02}$ \\
            \bottomrule
        \end{tabular}
        
        \label{tab:flux_MUTE}
    \end{table}

The systematic uncertainty includes a $10$\% relative uncertainty due to the spread in flux values across different configurations and a $3$\% relative uncertainty due to the seasonal modulation of muons. The statistical uncertainty is negligible and is ignored in the final flux value.
Taking the average between the maximum and minimum values from Tab.~\ref{tab:flux_MUTE}, our estimate of the cosmic-ray muon flux in the SNOLAB Cube Hall with MUTE software is:

\begin{equation*}
 \label{eq:MuonFlux_MUTE}
 \Phi_{\mathrm{MUTE}} = (4.37 \pm 0.45_{\textrm{sys}}) \times 10^{-10}\> \mu/\textrm{cm}^2/\textrm{s}.
\end{equation*}


\section{Standalone water tank measurement}
\label{sec:WaterTankMuonFlux} 

The muon flux measurement equation is given as:

\begin{equation}
	\label{eq:muon_flux_equation}
	\Phi_{\mu} = \frac{N_{\mu}}{A_{\textrm{eff}} \cdot t_{\textrm{live}} \cdot \varepsilon}   
\end{equation}

\noindent where $N_{\mu}$ is the number of muon candidates in the dataset, $A_{\textrm{eff}}$ is the effective area of the detector, $\varepsilon$ is the acceptance of event selection criteria, and $t_{\text{live}}$ is the summed detector livetime of the selected data collection runs (see Section~\ref{subsec:DEAP_data}).

The event selection for muon candidate events in the standalone water tank analysis requires 
that at least 30 MV PMTs register a signal,
that at least 70\% of the active MV PMTs register a signal,
that the summed signal from all MV PMTs is at least 250 photoelectrons,
and that the maximum fractional charge seen by any MV PMT is between 5\% and 40\% of the total charge.
To reduce instrumental backgrounds,
it is required that the number of MV PMTs registering a signal is at most the number of active MV PMTs,
and that the derivative of the summed waveform in the leading edge is less than $-1.0\,\textrm{ADC/ns}$ (i.e. that the negative slope is steeper than this value).
Further, it is required that
the integrated charge in the summed waveform above baseline is less than $1000\,\textrm{ADC}\cdot\textrm{ns}$, 
and that the peak of the summed waveform is within the expected 400~ns timing window.

\begin{figure}
    \centering
    \includegraphics[width=\linewidth]{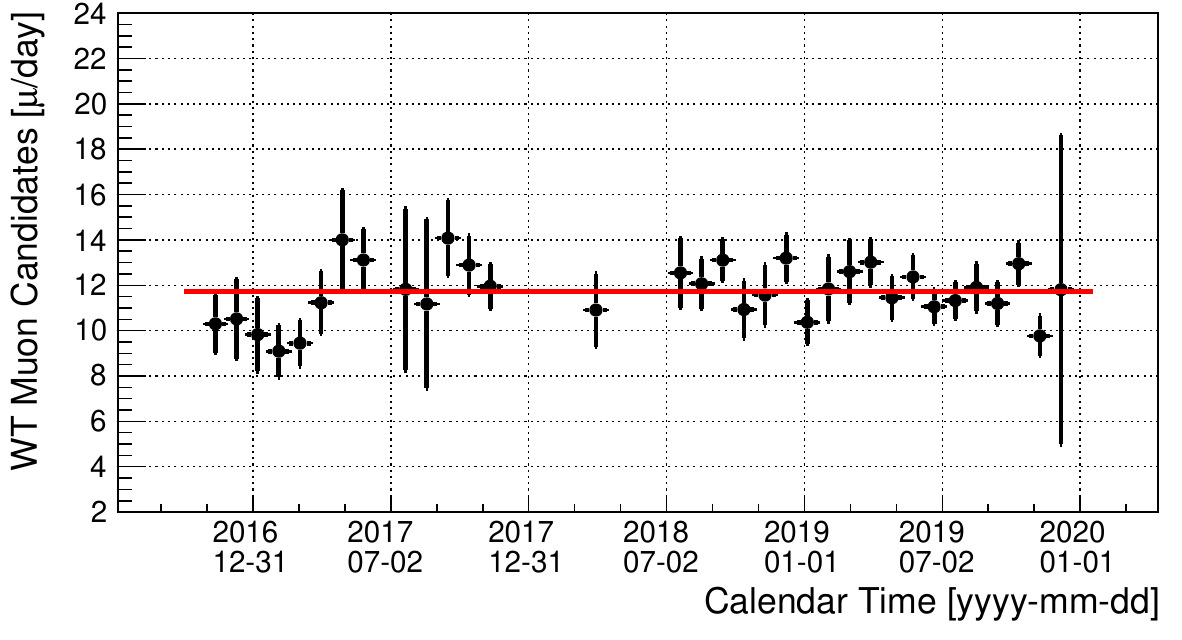}
    \caption{Observed rate of standalone water tank muon candidate events passing the selection criteria, taking into account the detector livetime in each bin. The calendar time is divided into 4-week bins, and the horizontal line represents the observed average rate.
    Bins with zero livetime are not shown.}
    \label{fig:RateOfWTMuons}
\end{figure}

Using these criteria, a total of $N_{\mu}^{\textrm{WT}} = 3309$ muon candidates are selected for the standalone water tank muon flux calculation. 
Fig.~\ref{fig:RateOfWTMuons} shows the corresponding event rate as a function of time.

To calculate the effective area of the water tank, a toy MC simulation is performed: \mbox{$2 \times 10^{11}$} downward-going muons are generated from a large plane above the simulated detector. The flux of upward-going muons is comparatively negligible and is ignored in this calculation. The angular distribution of muons is taken from the Mei--Hime model or the SNO model, described respectively by Eq.~\ref{eq:mei-hime_int_flux_slant_depth_zenith} and Eq.~\ref{eq:SNO_int_flux_zenith_dependent}, with the resulting difference in effective area taken as a systematic uncertainty.
Another source of systematic uncertainty comes from possible variations in the vertical water level over the data-taking period.
The effective area of the water tank is calculated based on the fraction of simulated muons that go through the water tank:
it is determined to be $A_{\textrm{eff}}^{\textrm{WT}}~=~(65.86~\pm~0.04_{\textrm{stat}}~\pm~0.31_{\textrm{sys}})~\textrm{m}^2$.

\begin{figure}
    \centering
    \includegraphics[width=0.96\linewidth]{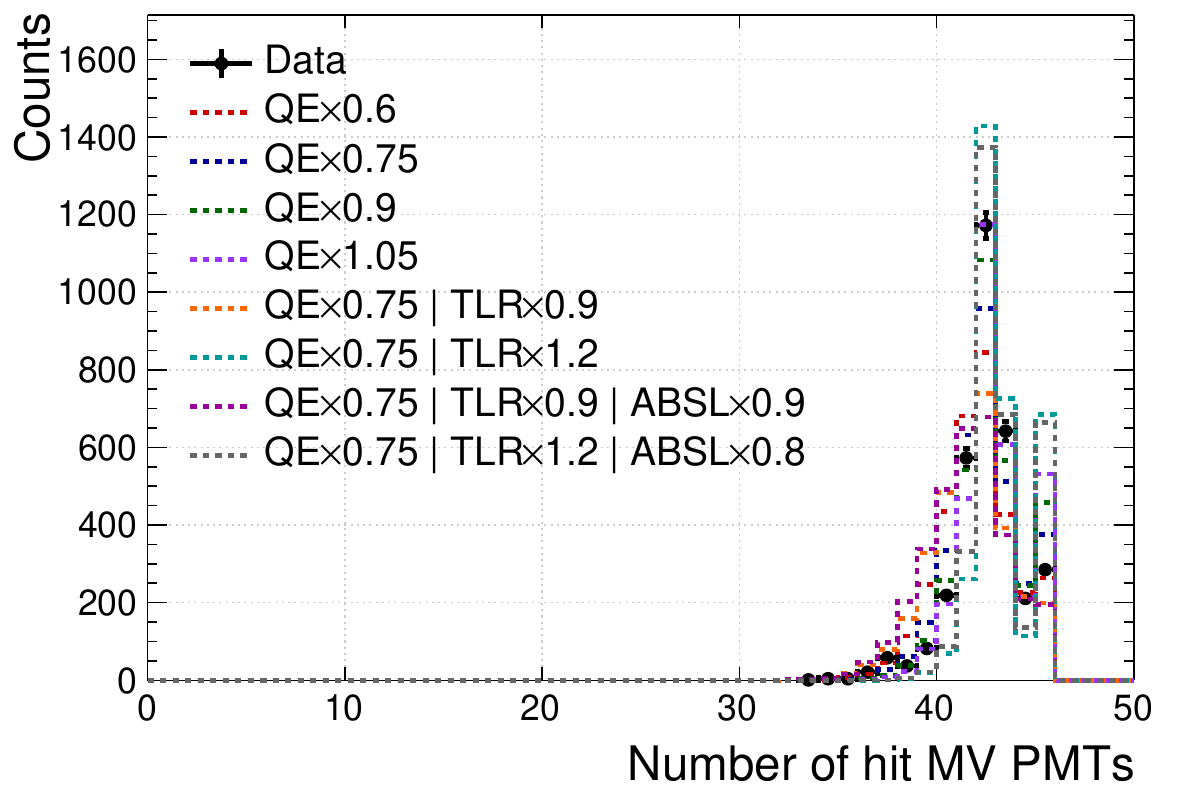}\\
    \includegraphics[width=0.96\linewidth]{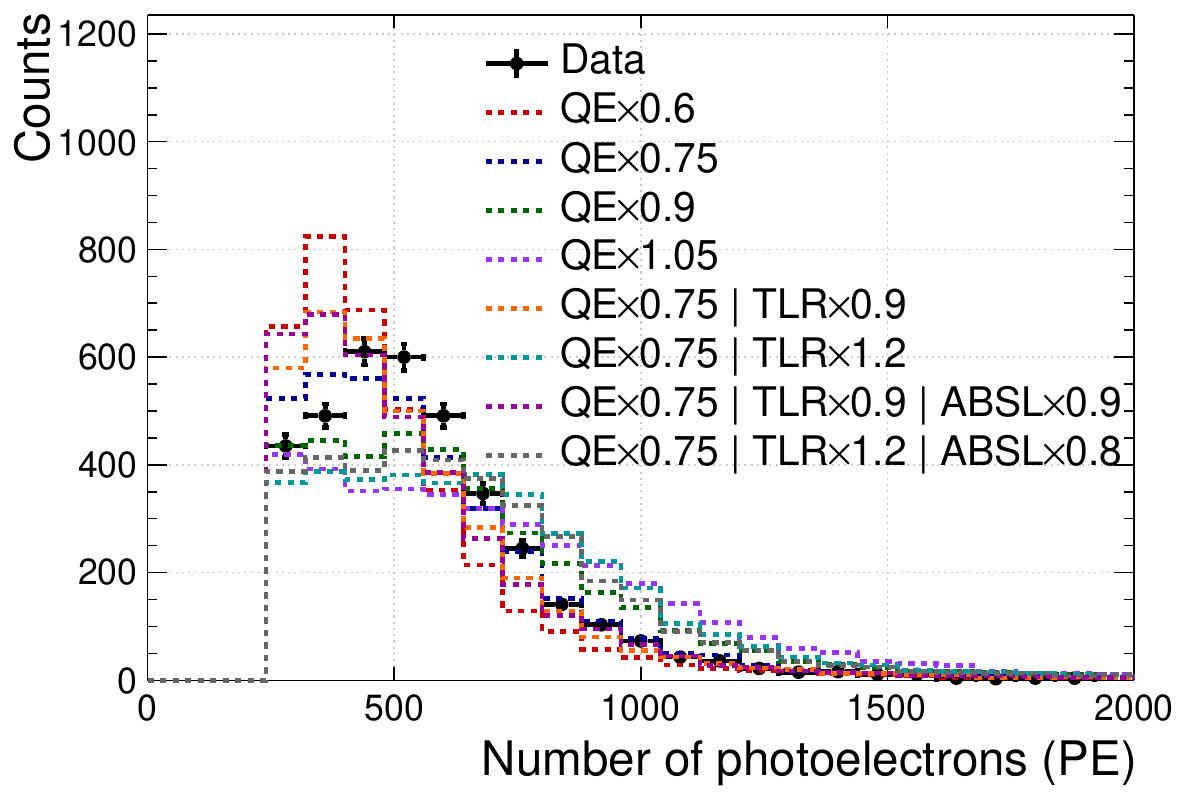}\\
    \includegraphics[width=0.96\linewidth]{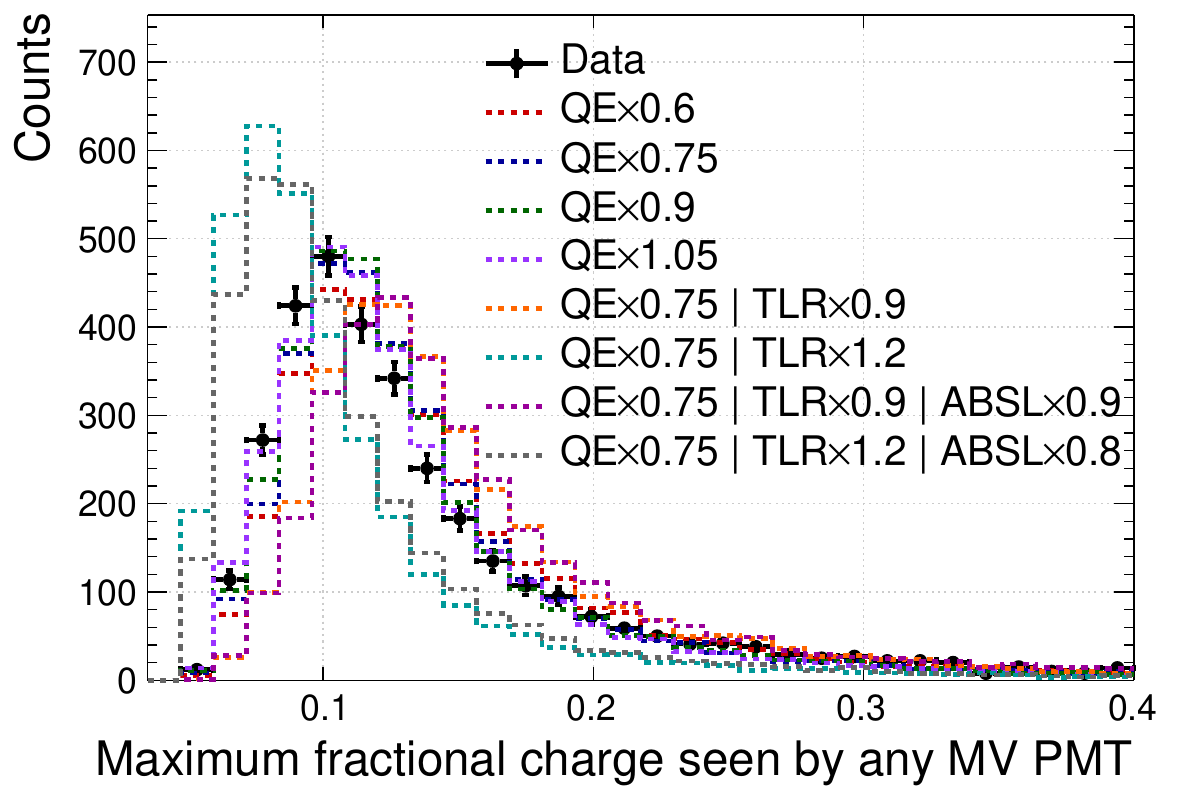}\\
    \caption{Data compared with various MC simulations consisting of different parameter configurations over several MV observables. The parameters considered are the quantum efficiency of the MV PMTs (QE), the water tank liner reflectivity (TLR), and the absorption length of photons in water (ABSL). These parameters are scaled from their nominal values as shown in the legend of the histograms.}
    \label{fig:Data-MC_match}
\end{figure}

The acceptance is calculated as the fraction of muon events passing the selection criteria out of the total number of muons traversing the detector.
It is factorized as $\varepsilon^{\textrm{WT}} = \varepsilon_1 \, \varepsilon_2$, 
where $\varepsilon_1$ accounts for efficiencies related to event reconstruction
and $\varepsilon_2$ is a data-driven acceptance accounting for events with overshoot in the summed event waveform and noise from inactive PMTs.
The acceptance term based on event reconstruction variables is evaluated using MC simulations carried out in \RAT. 
MC samples are generated based on three separate detector configurations based on the number of active MV PMTs, and the acceptance of simulated muons is taken as the weighted average over these three samples based on the livetime of the corresponding detector configuration in data.
Systematic uncertainties on this quantity are propagated from variations of the detector optical model, including the water tank liner reflectivity, the MV PMT quantum efficiency, and photon absorption length in water~\cite{smith1981optical}. 
Fig.~\ref{fig:Data-MC_match} shows a comparison of data and MC simulations with a range of parameter values in three key MV detector observables.
This MC-based acceptance term is estimated to be $\varepsilon_1 = (53~\pm~10_{\textrm{sys}})~\%$.

The data-driven acceptance term accounts for the two detector effects mentioned above that are not modeled by the \RAT\ simulations.
The acceptance for these selection criteria is calculated using the coincidence muon dataset by taking the fraction of events passing these selection criteria, giving $\varepsilon_2 = (88~\pm~2_{\textrm{stat}})~\%$.

The acceptance for selecting muons traversing the water tank is therefore $\varepsilon^{\textrm{WT}}~=~(46.6~\pm~1.1_{\textrm{stat}}~\pm~8.8_{\textrm{sys}})~\%$.

Thus, using Eq.~\ref{eq:muon_flux_equation} and propagating uncertainties, including a statistical component taken as the square root of the number of counts, the muon flux measured using the DEAP-3600 water tank data is

\begin{equation*}
	\label{eq:muon_flux_result_watertank}
	\Phi_{\mu}^{\textrm{WT}} = (4.42 \pm 0.13_{\textrm{stat}} \pm 0.84_{\textrm{sys}}) \times 10^{-10} \,\mu/\textrm{cm}^2/\textrm{s}.
\end{equation*}


\section{Coincidence Measurement}
\label{sec:CoincidenceMuonFlux}

To select coincident muon candidates, events observed in the MV are matched with events in the LAr detector using a $1\,\mu$s coincidence timing window. 
Additional selection criteria on water tank and LAr data are applied separately to select coincident muons. 
For the water tank data, looser criteria than in the standalone measurement are applied: here it is required that at least 3 MV PMTs register a signal, that the summed signal from all MV PMTs is at least 10 photoelectrons, 
and that the maximum fractional charge seen by any MV PMT is less than 50\% of the total charge.
Since muons interact with the LAr electromagnetically, they produce a lot of scintillation light read out by LAr \PMTs.
From the LAr detector, a threshold of 10$^5$ photoelectrons is applied; along with the coincidence timing requirement, this is sufficient to remove all background events. 

\begin{figure}[htb]
    \centering
    \includegraphics[width=\linewidth]{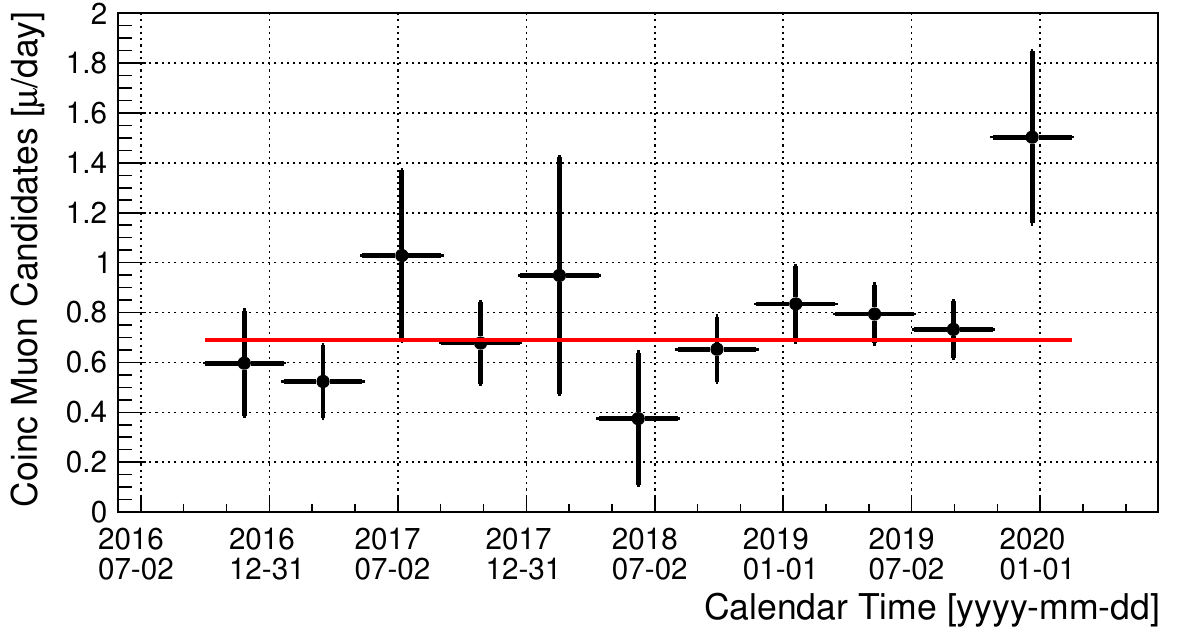}
    \caption{Observed rate of coincidence muon candidate events passing the selection criteria, taking into account the detector livetime in each bin. The calendar time is divided into 16-week bins, and the horizontal line represents the observed average rate.}
    \label{fig:RateOfCoinMuons}
\end{figure}

A total of $N_{\mu}^{\textrm{coinc}} = 227$ muon candidates pass the selection criteria for the coincidence muon flux calculation.
Out of these, 167 events are also counted in $N_{\mu}^{\textrm{WT}}$, so the two measurements are correlated.
Fig.~\ref{fig:RateOfCoinMuons} shows the rate of coincidence muon candidate events as a function of time.

\begin{figure}[htb]
    \centering
    \includegraphics[width=\linewidth]{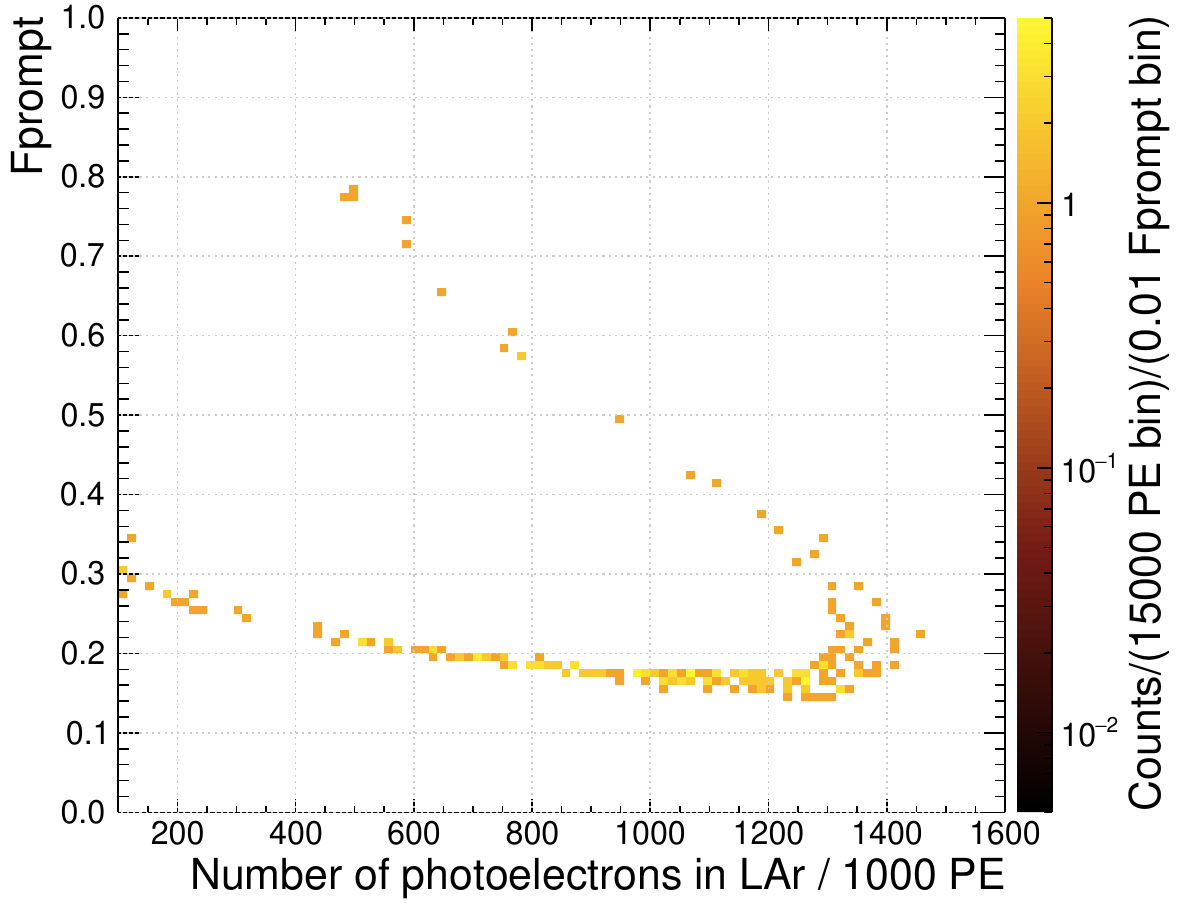}
    \caption{The PSD observable Fprompt with a 150~ns prompt time window~\cite{deap_psd_2021} versus the number of photoelectrons observed in the LAr detector for coincidence muon candidate events.
    With increasing energy deposited into the LAr volume, the effect of PMT saturation and digitizer clipping is visible as the prompt peak is not fully recorded, resulting in a trend where Fprompt values curve from 0.3 (at $10^5$ photoelectrons) down to 0.15.  
    Without correcting for these effects, the limit of the DEAP-3600 detector is reached around $1.5\times10^6$ photoelectrons, beyond which the trend curves upward in Fprompt for very high-energy muon events that highly saturate the inner PMTs, leading to an overall decrease in the number of photoelectrons recorded.}
    \label{fig:Fprompt_vs_qPE}
\end{figure}


For the effective area calculation,
even though the LAr detector of DEAP-3600 in the partial fill configuration is not a perfect sphere,
due to the flat overburden over SNOLAB and supported by the angular distribution models of Eq.~\ref{eq:mei-hime_int_flux_slant_depth_zenith} and Eq.~\ref{eq:SNO_int_flux_zenith_dependent},
the vast majority of incoming muons arrive within \mbox{$\theta < 0.9$}~rad. 
Therefore, the effective area of the LAr volume seen by incoming muons is that of a disk of radius \mbox{$(845.6\pm0.9)$}~mm~\cite{adhikari2023precision}, i.e.
$A_{\textrm{eff}}^{\textrm{coinc}}=(2.246\pm0.005_{\textrm{sys}})~\textrm{m}^2$.
This geometrical argument was cross-checked using MC simulations of coincidence muon events.


In the coincidence analysis, the acceptance of the event selection criteria applied to data from the MV and the LAr detector are calculated separately, factorizing as $\varepsilon^{\textrm{coinc}} = \varepsilon_3 \, \varepsilon_4$. 
The simulated muon samples described in Section~\ref{sec:WaterTankMuonFlux}\ are used to calculate the cut acceptance of the aforementioned looser selection criteria applied to MV data, yielding $\varepsilon_3 = (95.8 \pm 0.6_{\textrm{stat}} \pm 2.3_{\textrm{sys}} )~\%$,
where variations in the detector optical model used in simulations are the dominant systematic uncertainty.

The acceptance of the LAr scintillation threshold is evaluated by calibrating the energy response of simulated muons against observed data using a quadratic energy-response model specific to very high-energy events.
Taken into account in this model is the observed decrease in the number of photoelectrons detected when the energy deposited in LAr is very high, due to PMT saturation and digitizer clipping: 
this effect is displayed in Fig.~\ref{fig:Fprompt_vs_qPE}.
The acceptance calculation yields that a fraction $\varepsilon_4 = (99.90 \pm 0.02_{\textrm{stat}} \pm 0.03_{\textrm{sys}} )~\%$ of simulated muons crossing the LAr detector deposit sufficient energy to pass this requirement.
Indeed, a calculation from the scintillation light yield shows that the threshold is expected to be surpassed by all muons traversing at least 7~cm in the LAr. 
The systematic uncertainty on this component of the acceptance is propagated from the energy response model parameters. 

Therefore, the overall acceptance for selecting muons traversing both the MV and the LAr detector is found to be $\varepsilon^{\textrm{coinc}}~=~(95.7~\pm~0.6_{\textrm{stat}}~\pm~2.3_{\textrm{sys}})~\%$.


Using Eq.~\ref{eq:muon_flux_equation} and propagating uncertainties, the measured muon flux in the coincidence analysis is

\begin{equation*}
    \label{eq:coinMuon_result}
    \Phi_{\mu}^{\textrm{coinc}} = (3.71 \pm 0.25_{\textrm{stat}} \pm 0.09_{\textrm{sys}}) \times 10^{-10} \,\mu/\textrm{cm}^2/\textrm{s}.
\end{equation*}


\section{Results and conclusion}
\label{sec:results}

This work presents a direct and independent measurement of the muon flux at SNOLAB with the data collected in 2016-2020 by the DEAP-3600 experiment. 
Two measurements are performed, one strictly using events observed in the MV detector (the standalone water tank measurement) and the other using events observed in coincidence between the MV and the LAr detector. 
In each case, the muon flux is determined in a cut-and-count analysis taking as input the number of observed events, the detector livetime and effective area, and the event-selection acceptance.
These measurements are correlated, as there is overlap in the muon candidates passing each event selection.

\begin{table}[htb]
    \centering
    \renewcommand*{\arraystretch}{1.15}
    \setlength{\tabcolsep}{2.5pt}
    \caption{Summary of the DEAP-3600 standalone water tank and coincidence measurements of the muon flux.}
    \vspace{0.25cm}
    \label{tab:Muon_flux_summary}
    \begin{tabular}{l c c}
        \toprule
         & \textbf{Standalone water tank} & \textbf{Coincidence} \\
        \midrule
        $N_\mu$   & 3309    & 227  \\
        $t_{\textrm{live}}$ [s]   & $2.44\times10^7$              & $2.85\times10^7$ \\
        $A_\text{eff}$ [m$^2$]     & $65.86 \pm 0.04_{\textrm{stat}} \pm 0.31_{\textrm{sys}}$   & $2.246 \pm 0.005_{\textrm{sys}}$ \\
        $\varepsilon$ [\%]         & $46.6 \pm 1.1_{\textrm{stat}} \pm 8.8_{\textrm{sys}}$      & $95.7\pm0.6_{\textrm{stat}}\pm2.3_{\textrm{sys}}$ \\
        \midrule
        $\Phi_\mu\ [\times 10^{-10}$                 & $4.42 \pm 0.13_{\textrm{stat}} \pm 0.84_{\textrm{sys}}$    & $3.71 \pm 0.25_{\textrm{stat}} \pm 0.09_{\textrm{sys}}$ \\
        \phantom{$\Phi_\mu$\ \,}$\mu/\text{cm}^2/\text{s}]$ & & \\
        \bottomrule
    \end{tabular}
\end{table}

\begin{figure}[htb!]
    \centering
    \includegraphics[width=\columnwidth]{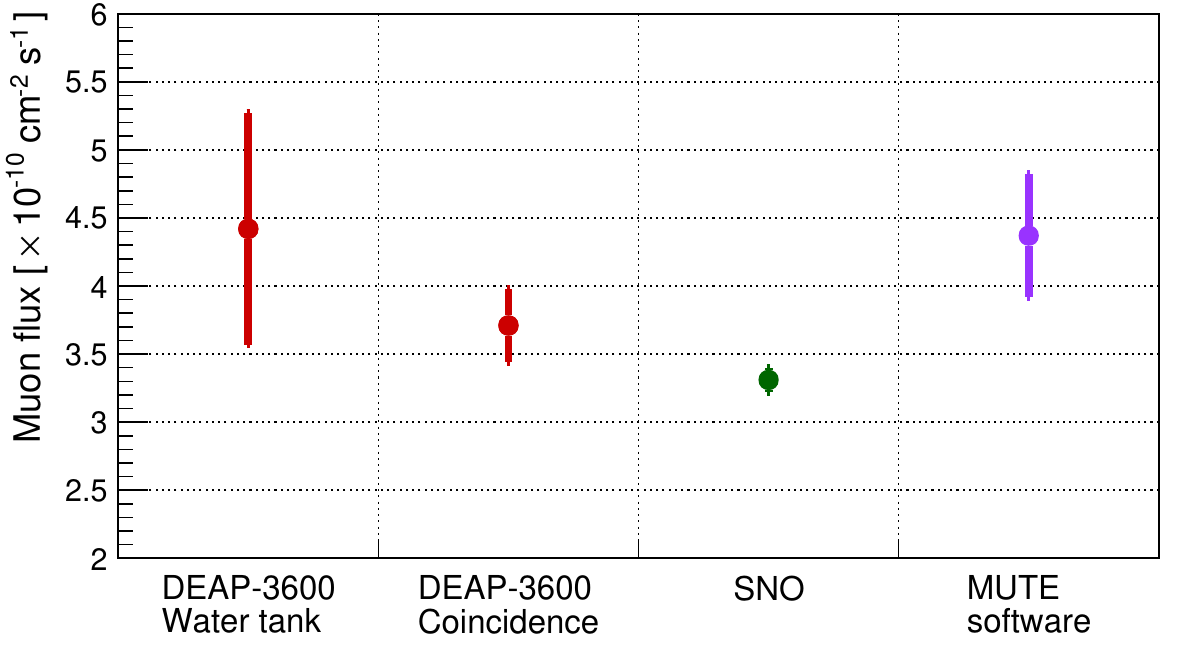}
    \caption{Comparison of the muon flux measured in this work with the cosmic-ray muon flux measurement by SNO and the estimate with MUTE software.}
    \label{fig:Flux_results}
\end{figure}

The results are summarized in Tab.~\ref{tab:Muon_flux_summary}.
The two DEAP-3600 measurements are mutually consistent within their combined uncertainties.
Fig.~\ref{fig:Flux_results} shows a comparison with the cosmic-ray muon flux measurement by SNO and the estimate using MUTE software. 
The upward-going neutrino-induced component of the muon flux, contributing $<1\%$ of the total~\cite{SNO_muonflux}, is included in the DEAP-3600 measurements but not in the SNO measurement and MUTE calculations.
Without correcting for this small effect, the standalone water tank muon flux measurement differs from the SNO measurement by $1.3\sigma$ (where $\sigma$ is the sum in quadrature of statistical and systematic uncertainties from both DEAP-3600 and SNO), while the coincidence measurement differs by $1.4\sigma$. 
The estimate of the cosmic-ray muon flux with MUTE software is also consistent with the DEAP-3600 results, and differs from the SNO measurement by $2.3\sigma$.

The uncertainties from this work are larger than in the SNO result due to DEAP-3600's smaller detector size, reduced PMT coverage in the MV, detector instabilities during data taking, and relatively low light-collection efficiency in the MV. 
In the standalone water tank measurement, the dominant systematic uncertainty comes from the component of the acceptance estimated using MC simulations, due to uncertainties in optical model parameters of the MV detector.  
In the coincidence measurement, the systematic uncertainty is significantly smaller, because requiring muons to traverse both detectors suppresses the impact of instrumental effects and the acceptance of the event selection for muons crossing the LAr detector is very high.
The coincidence measurement of the muon flux is limited by its statistical uncertainty.

An important background source for dark matter searches in underground experiments arises from the production of cosmogenic neutrons due to muon interactions with the surrounding rock and detector material. 
Therefore, an accurate determination of the muon flux is essential for background estimation, improving the sensitivity of current and future rare-event search experiments.
In future work, an improved muon reconstruction and tracking algorithm could enable a more precise measurement of muons and muon-induced background events.

In conclusion, the muon flux measurements performed in this analysis are consistent with the previous SNO measurement and with the MUTE software calculation. 
These measurements by the DEAP-3600 experiment serve as an important validation of the muon rate observed in Cube Hall at SNOLAB.
\\


\section{Acknowledgements}
\label{sec:Acknowledgements}

We thank the Natural Sciences and Engineering Research Council of Canada (NSERC),
the Canada Foundation for Innovation (CFI),
the Ontario Ministry of Research and Innovation (MRI), 
and Alberta Advanced Education and Technology (ASRIP),
the University of Alberta,
Carleton University, 
Queen's University,
the Canada First Research Excellence Fund through the Arthur B.~McDonald Canadian Astroparticle Physics Research Institute,
SECIHTI Project No. CBF-2025-I-1589,
DGAPA UNAM Grant No. PAPIIT IN102326,
the European Research Council Project (ERC StG 279980),
the UK Science and Technology Facilities Council (STFC) (ST/K002570/1 and ST/R002908/1),
the Leverhulme Trust (ECF-20130496),
the Russian Science Foundation (Grant No. 21-72-10065),
the Spanish Ministry of Science and Innovation (PID2022-138357NB-C22),
the National Science Centre, Poland (2022/47/B/ST2/02015),
the International Research Agenda Programmes of the Foundation for Polish Science: AstroCeNT (MAB/2018/7), funded from the European Regional Development Fund, 
and Astrocent (FENG.02.01-IP.05-A015/25), co-financed by the European Union under the European Funds for Smart Economy 2021-2027 (FENG); 
Teaming for Excellence grant Astrocent Plus (101137080) funded by the European Union with complementary national funding from the Polish Ministry of Science and Higher Education (MNiSW/2025/DIR/811).
Studentship support from
the Rutherford Appleton Laboratory Particle Physics Division,
STFC and SEPNet PhD is acknowledged.
We thank SNOLAB and its staff for support through underground space, logistical, and technical services.
SNOLAB operations are supported by the CFI
and the Province of Ontario MRI,
with underground access provided by Vale at the Creighton mine site.
We thank Vale for their continuing support, including the work of shipping the acrylic vessel underground.
We gratefully acknowledge the support of the Digital Research Alliance of Canada,
Calcul Qu\'ebec,
the Centre for Advanced Computing at Queen's University,
the Computational Centre for Particle and Astrophysics (C2PAP) at the Leibniz Supercomputer Centre (LRZ)
and SNOLAB
for providing the computing resources required to undertake this work.

\bibliography{main}

@article{das2026rare,
  title={Rare event searches using cryogenic detectors via direct detection methods},
  author={Das, S and others}, 
  journal={Modern Phys. Lett. A},
  volume={41},
  number={12n13},
  pages={2630002},
  year={2026},
  publisher={World Scientific}
}

@article{pvevc2024muon,
  title={Muon-induced background in a next-generation dark matter experiment based on liquid xenon},
  author={P{\v{e}}{\v{c}}, Viktor and others}, 
  journal={Eur. Phys. J. C},
  volume={84},
  number={5},
  pages={481},
  year={2024},
  publisher={Springer}
}

@article{abreu2026cosmogenic,
  title={Cosmogenic neutron production in water at {SNO+}},
  author={Abreu, M and others},
  collaboration={{SNO+} Collaboration},
  journal={Phys. Rev. D},
  volume={113},
  number={5},
  pages={052014},
  year={2026},
  publisher={APS}
}

@article{xia2026progress,
  title={Progress and prospects in the underground laboratories' search for dark matter},
  author={Xia, Qing and Canonica, Lucia},
  journal={Communications Physics},
  volume={9},
  number={1},
  pages={105},
  year={2026},
  publisher={Nature Publishing Group UK London}
}

@article{abi2021prospects,
  title={Prospects for beyond the {Standard Model} physics searches at the deep underground neutrino experiment},
  author={Abi, B and others},
  collaboration={{DUNE} Collaboration},
  journal={Eur. Phys. J. C},
  volume={81},
  number={4},
  pages={322},
  year={2021},
  publisher={Springer}
}

@article{aprile2025neutron,
  title={The neutron veto of the {XENONnT} experiment: results with demineralized water},
  author={Aprile, Elena and others},
  collaboration={{XENONnT} Collaboration},
  journal={Eur. Phys. J. C},
  volume={85},
  number={6},
  pages={695},
  year={2025},
  publisher={Springer}
}

@article{angloher2024water,
  title={Water Cherenkov muon veto for the {COSINUS} experiment: design and simulation optimization},
  author={Angloher, G and others},
  collaboration={{COSINUS} Collaboration},
  journal={Eur. Phys. J. C},
  volume={84},
  number={5},
  pages={551},
  year={2024},
  publisher={Springer}
}

@article{burlac2025early,
  title={Early results of the {LEGEND-200} experiment},
  author={Burlac, Nina and others},
  collaboration={{LEGEND} Collaboration},
  journal={Nucl. Instrum. Methods Phys. Res. A},
  volume={1080},
  pages={170779},
  year={2025},
  publisher={Elsevier}
}

@article{diamond2025community,
  title={Community Report from the 2025 {SNOLAB} Future Projects Workshop},
  author={Diamond, M. D. and others},
  journal={arXiv preprint},
  eprint = {2507.11368},
  archivePrefix = {arXiv},
  primaryClass={hep-ex},
  url={https://arxiv.org/abs/2507.11368}, 
  year={2025},

}

@article{fedynitch2019hadronic,
  title={Hadronic interaction model {SIBYLL 2.3c} and inclusive lepton fluxes},
  author={Fedynitch, Anatoli and others},
  journal={Phys. Rev. D},
  volume={100},
  number={10},
  pages={103018},
  year={2019},
  publisher={APS}
}

@article{MUTE,
	author = {Fedynitch, Anatoli and Woodley, William and Piro, Marie-Cecile},
	title = {On the Accuracy of Underground Muon Intensity Calculations},
	doi = {10.3847/1538-4357/ac5027},
	journal = {Astrophys. J.},
	volume = {928},
	number = {1},
	pages = {27},
	year = {2022}
}

@misc{MUTE_software,
  author       = {William Woodley and Anatoli Fedynitch},
  title        = {wjwoodley/mute: {MUTE} 2.0.0},
  month        = jul,
  year         = 2022,
  publisher    = {Zenodo},
  version      = {2.0.0},
  doi          = {10.5281/zenodo.6841971},
  url          = {https://doi.org/10.5281/zenodo.6841971},
}

@article{PROPOSAL,
	title = {{PROPOSAL}: A tool for propagation of charged leptons},
	journal = {Computer Physics Communications},
	volume = {184},
	number = {9},
	pages = {2070--2090},
	year = {2013},
	issn = {0010-4655},
	doi = {10.1016/j.cpc.2013.04.001},
	url = {https://www.sciencedirect.com/science/article/pii/S0010465513001355},
	author = {Koehne, J.-H. and others}
}

@article{MCEq,
	author = {Fedynitch, Anatoli and others},
	editor = {Berge, D. and de Roeck, A. and Mangano, M. and Pattison, B.},
	title = {{Calculation of conventional and prompt lepton fluxes at very high energy}},
	eprint = {1503.00544},
	archivePrefix = {arXiv},
	primaryClass = {hep-ph},
	doi = {10.1051/epjconf/20159908001},
	journal = {EPJ Web Conf.},
	volume = {99},
	pages = {08001},
	year = {2015}
}

@article{H3a,
	author = {Gaisser, Thomas K. and others},
	title = {{Cosmic Ray Energy Spectrum from Measurements of Air Showers}},
	eprint = {1303.3565},
	archivePrefix = {arXiv},
	primaryClass = {astro-ph.HE},
	doi = {10.1007/s11467-013-0319-7},
	journal = {Front. Phys. (Beijing)},
	volume = {8},
	pages = {748--758},
	year = {2013}
}

@article{GSF,
	author = {Dembinski, Hans Peter and others},
	title = {{Data-driven model of the cosmic-ray flux and mass composition from 10 GeV to $10^{11}$ GeV}},
	eprint = {1711.11432},
	archivePrefix = {arXiv},
	primaryClass = {astro-ph.HE},
	doi = {10.22323/1.301.0533},
	journal = {PoS},
	volume = {ICRC2017},
	pages = {533},
	year = {2018}
}

@article{Geant4_1,
	title = {Recent developments in {Geant4}},
	journal = {Nucl. Instrum. Methods Phys. Res. A},
	volume = {835},
	pages = {186--225},
	year = {2016},
	issn = {0168-9002},
	doi = {10.1016/j.nima.2016.06.125},
	url = {https://www.sciencedirect.com/science/article/pii/S0168900216306957},
	author = {Allison, J. and others}
}

@article{Geant4_2,
	author = {Allison, J. and others},
	journal = {IEEE Transactions on Nuclear Science},
	title = {{Geant4} developments and applications},
	year = {2006},
	volume = {53},
	number = {1},
	pages = {270--278},
	doi = {10.1109/TNS.2006.869826}
}

@book{bass_handbook_2009,
	series = {Handbook of {Optics}},
	title = {Handbook of {Optics}, {Third} {Edition} {Volume} {II}: {Design}, {Fabrication} and {Testing}, {Sources} and {Detectors}, {Radiometry} and {Photometry}},
	isbn = {978-0-07-162927-0},
	url = {https://books.google.com/books?id=w3zD5LwGu0sC},
	publisher = {McGraw-Hill Education},
	author = {Bass, M. and others},
	edition = {3},
	year = {2009},
	lccn = {2009036597}
}

@article{SNO_muonflux,
	author = {Aharmim, B. and others},
	collaboration = {SNO Collaboration},
	title = {Measurement of the Cosmic Ray and Neutrino-Induced Muon Flux at the {Sudbury Neutrino Observatory}},
	doi = {10.1103/PhysRevD.80.012001},
	journal = {Phys. Rev. D},
	volume = {80},
	pages = {012001},
	year = {2009}
}

@article{ROOT_NIMA_1997,
    author = "Brun, R. and Rademakers, F.",
    editor = "Werlen, M. and Perret-Gallix, D.",
    title = "{ROOT: An object oriented data analysis framework}",
    doi = "10.1016/S0168-9002(97)00048-X",
    journal = "Nucl. Instrum. Meth. A",
    volume = "389",
    pages = "81--86",
    year = "1997"
}

@article{ajaj2019search,
  title={Search for dark matter with a 231-day exposure of liquid argon using {DEAP-3600} at {SNOLAB}},
  author={Ajaj, R and Amaudruz, P-A and Araujo, GR and Baldwin, M and Batygov, M and Beltran, B and Bina, CE and Bonatt, J and Boulay, MG and Broerman, B and others},
  journal={Phys. Rev. D},
  volume={100},
  number={2},
  pages={022004},
  year={2019},
  publisher={APS}
}

@phdthesis{AndrewThesis,
    author = "Erlandson, Andrew",
    title = "{First observation of solar neutrino absorption on \textsuperscript{40}Ar using the DEAP-3600 detector}",
    doi = "10.22215/etd/2024-16484",
    school = "Carleton U.",
    year = "2024"
}

@article{mei_muon-induced_2006,
	title = {Muon-induced background study for underground laboratories},
	volume = {73},
	url = {https://link.aps.org/doi/10.1103/PhysRevD.73.053004},
	doi = {10.1103/PhysRevD.73.053004},
	number = {5},
	journal = {Phys. Rev. D},
	author = {Mei, D.-M. and Hime, A.},
	month = mar,
	year = {2006},
	pages = {053004}
}

@article{bellini_cosmogenic_2013,
	title = {Cosmogenic backgrounds in {Borexino} at 3800 m water-equivalent depth},
	volume = {2013},
	url = {http://stacks.iop.org/1475-7516/2013/i=08/a=049},
	doi = {10.1088/1475-7516/2013/08/049},
	number = {08},
	journal = {J. Cosmol. Astropart. Phys.},
	author = {Bellini, G. and others},
	collaboration = {{Borexino Collaboration}},
	year = {2013},
	pages = {049}
}

@article{empl_fluka_2014,
	title = {A {FLUKA} Study of {$\beta$}-delayed Neutron Emission for the Ton-size {DarkSide} Dark Matter Detector},
	url = {http://arxiv.org/abs/1407.6628},
	journal = {arXiv:1407.6628},
	author = {Empl, Anton and Hungerford, Ed V.},
	month = jul,
	year = {2014}
}

@article{kamland_collaboration_production_2010,
	title = {Production of radioactive isotopes through cosmic muon spallation in {KamLAND}},
	volume = {81},
	url = {http://link.aps.org/doi/10.1103/PhysRevC.81.025807},
	doi = {10.1103/PhysRevC.81.025807},
	number = {2},
	journal = {Phys. Rev. C},
	collaboration = {{KamLAND Collaboration}},
	author = {Abe, S. and others},
	month = feb,
	year = {2010},
	pages = {025807}
}

@article{amaudruz_design_2019,
	title = {Design and construction of the {DEAP}-3600 dark matter detector},
	volume = {108},
	url = {http://www.sciencedirect.com/science/article/pii/S0927650518300914},
	doi = {10.1016/j.astropartphys.2018.11.007},
	journal = {Astropart. Phys.},
	collaboration = {DEAP Collaboration},
	author = {Amaudruz, P. -A. and others},
	month = mar,
	year = {2019},
	pages = {1--23}
}

@article{AMAUDRUZ2019373,
title = {In-situ characterization of the Hamamatsu R5912-HQE photomultiplier tubes used in the DEAP-3600 experiment},
journal = {Nucl. Instrum. Methods Phys. Res. A},
volume = {922},
pages = {373-384},
year = {2019},
issn = {0168-9002},
doi = {https://doi.org/10.1016/j.nima.2018.12.058},
url = {https://www.sciencedirect.com/science/article/pii/S0168900218318552},
author = {P.-A. Amaudruz and others},
collaboration = {DEAP Collaboration},
}

@article{pierog2015epos,
  title={{EPOS LHC}: Test of collective hadronization with data measured at the {CERN Large Hadron Collider}},
  author={Pierog, Tanguy and others},
  journal={Phys. Rev. C},
  volume={92},
  number={3},
  pages={034906},
  year={2015},
  publisher={APS}
}

@inproceedings{hall2020snolab,
	author = {Hall, J.},
	title = {The {SNOLAB} underground laboratory},
	booktitle = {Journal of Physics: Conference Series},
	volume = {1468},
	pages = {012252},
	year = {2020},
	organization = {IOP Publishing}
}

@article{deap_psd_2021,
	title = {Pulse-shape discrimination against low-energy {Ar}-39 beta decays in liquid argon with 4.5 tonne-years of {DEAP-3600} data},
	author = {Adhikari, P. and others},
	journal = {Eur. Phys. J. C},
	collaboration = {{DEAP Collaboration}},
	number = {9},
	pages = {823},
	volume = {81},
	year = {2021}
}

@article{adhikari2023precision,
  title={Precision measurement of the specific activity of {39Ar} in atmospheric argon with the {DEAP-3600} detector},
  author = {Adhikari, P. and others},
  collaboration={{DEAP collaboration}},
  journal={Eur. Phys. J. C},
  volume={83},
  number={7},
  pages={642},
  year={2023},
  publisher={Springer}
}

@article{malgin_phenomenology_2017,
	title = {Phenomenology of muon-induced neutron yield},
	volume = {96},
	url = {https://link.aps.org/doi/10.1103/PhysRevC.96.014605},
	doi = {10.1103/PhysRevC.96.014605},
	number = {1},
	journal = {Phys. Rev. C},
	author = {Malgin, A. S.},
	month = jul,
	year = {2017},
	pages = {014605}
}

@article{aglietta1998muon,
  title={Muon “depth-intensity” relation measured by the {LVD} underground experiment and cosmic-ray muon spectrum at sea level},
  author={Aglietta, M. and others},
  collaboration={LVD Collaboration},
  journal={Phys. Rev. D},
  volume={58},
  number={9},
  pages={092005},
  year={1998},
  publisher={APS}
}

@article{ambrosio1995vertical,
  title={Vertical muon intensity measured with {MACRO at the Gran Sasso laboratory}},
  author={Ambrosio, M. and others},
  collaboration={MACRO Collaboration},
  journal={Phys. Rev. D},
  volume={52},
  number={7},
  pages={3793},
  year={1995},
  publisher={APS}
}

@article{thunman1996charm,
	title = {Charm production and high energy atmospheric muon and neutrino fluxes},
	author = {Thunman, M. and Ingelman, G. and Gondolo, P.},
	journal = {Astroparticle Physics},
	volume = {5},
	number = {3-4},
	pages = {309--332},
	year = {1996},
	publisher = {Elsevier}
}

@article{smith1981optical,
  title={Optical properties of the clearest natural waters (200--800 nm)},
  author={Smith, Raymond C and Baker, Karen S},
  journal={Applied Optics},
  volume={20},
  number={2},
  pages={177--184},
  year={1981},
  publisher={Optical Society of America}
}

@article{ianni_review_2017,
	title = {Review of technical features in underground laboratories},
	volume = {32},
	url = {https://www-worldscientific-com.libproxy.mit.edu/doi/10.1142/S0217751X17430011},
	doi = {10.1142/S0217751X17430011},
	number = {30},
	journal = {International Journal of Modern Physics A},
	author = {Ianni, Aldo},
	month = oct,
	year = {2017},
	pages = {1743001}
}

@article{palusova_natural_2020,
	title = {Natural radionuclides as background sources in the {Modane} underground laboratory},
	volume = {216},
	url = {http://www.sciencedirect.com/science/article/pii/S0265931X19308525},
	doi = {10.1016/j.jenvrad.2020.106185},
	journal = {J. Environ. Radioact.},
	author = {Pal\u{s}ov\'a, V. and Breier, R. and Chauveau, E. and Piquemal, F. and Povinec, P. P.},
	month = may,
	year = {2020},
	pages = {106185}
}

@article{agnes_veto_2016,
  title = {The veto system of the {DarkSide}-50 experiment},
  author = {Agnes, P. and others},
  collaboration = {{DarkSide Collaboration}},
  journal = {J. Instrum.},
  volume = {11},
  number={03},
  pages = {P03016},
  year = {2016},
  month = mar,
  doi = {10.1088/1748-0221/11/03/P03016}
}

@article{an_muon_2015,
	title = {The muon system of the {Daya} {Bay} {Reactor} antineutrino experiment},
	volume = {773},
	url = {http://www.sciencedirect.com/science/article/pii/S0168900214011024},
	doi = {10.1016/j.nima.2014.09.070},
	journal = {Nucl. Instrum. Methods Phys. Res. A},
	collaboration = {{Daya Bay}},
	author = {An, F. P. and others},
	month = feb,
	year = {2015},
	pages = {8--20}
}

@article{calkins_prototyping_2015,
	title = {Prototyping an active neutron veto for {SuperCDMS}},
	volume = {1672},
	url = {https://aip.scitation.org/doi/abs/10.1063/1.4928018},
	doi = {10.1063/1.4928018},
	number = {1},
	journal = {AIP Conference Proceedings},
	author = {Calkins, Robert and Loer, Ben and Orrell, John L.},
	month = aug,
	year = {2015},
	pages = {140002}
}

@article{oconnell_muon-induced_1988,
	title = {Muon-induced radioactivity in underground detectors},
	volume = {38},
	url = {https://link.aps.org/doi/10.1103/PhysRevD.38.2277},
	doi = {10.1103/PhysRevD.38.2277},
	number = {7},
	journal = {Phys. Rev. D},
	author = {O’Connell, J. S. and Schima, F. J.},
	month = oct,
	year = {1988},
	pages = {2277--2279}
}

@article{carson_neutron_2004,
	title = {Neutron background in large-scale xenon detectors for dark matter searches},
	volume = {21},
	url = {http://www.sciencedirect.com/science/article/pii/S0927650504000751},
	doi = {10.1016/j.astropartphys.2004.05.001},
	number = {6},
	journal = {Astropart. Phys.},
	author = {Carson, M. J. and others},
	month = sep,
	year = {2004},
	pages = {667--687}
}

@inproceedings{ford2012snolab,
  title={{SNOLAB}: Review of the facility and experiments},
  author={Ford, Richard},
  booktitle={AIP Conference Proceedings},
  volume={1441},
  pages={521--524},
  year={2012},
  organization={American Institute of Physics}
}

\end{document}